# Differences in the Mechanical Properties of Monolayer and Multilayer WSe$_2$/MoSe$_2$

Y. M. Jaques[1,2,3], P. Manimunda[4], Y. Nakanishi[2], S. Susarla[2], C. F. Woellner[1,2,3], S. Bhowmick[4], S. A. S. Asif[4], and D. S. Galvão[1,3], C. S. Tiwary[2], and P. M. Ajayan[2]

[1]*Applied Physics Department, State University of Campinas, 13083-859 Campinas-SP, Brazil*

[2]*Materials Science and Nanoengineering, Rice University, Houston, Texas 77005, USA*

[3]*Center for Computational Engineering & Sciences, State University of Campinas, Campinas-SP, Brazil*

[4]*Bruker Nano Surfaces, Minneapolis, MN, USA*

ABSTRACT

*Transition metal dichalcogenides are 2D structures with remarkable electronic, chemical, optical and mechanical properties. Monolayer and crystal properties of these structures have been extensively investigated, but a detailed understanding of the properties of their few-layer structures are still missing. In this work we investigated the mechanical differences between monolayer and multilayer WSe2 and MoSe2, through fully atomistic molecular dynamics simulations (MD). It was observed that single layer WSe2/MoSe2 deposited on silicon substrates have larger friction coefficients than 2, 3 and 4 layered structures. For all considered cases it is always easier to peel off and/or to fracture MoSe$_2$ structures. These results suggest that the interactions between first layer and substrate are stronger than interlayer interactions themselves. Similar findings have been reported for other nanomaterials and it has been speculated whether this is a universal-like behavior for 2D layered materials. We have also analyzed fracture patterns. Our results show that fracture is chirality dependent with crack propagation preferentially perpendicular to W(Mo)-Se bonds and faster for zig-zag-like defects.*

**INTRODUCTION**

Two-dimensional (2D) semiconducting materials with atomic thickness have emerged as potential candidates to improve device energy efficiency with the possibility of being optically transparent and mechanically flexible [1-3]. One class of such materials is transition metal dichalcogenides (TMDC), which are structures of the type **MX$_2$**, where **M** is metal atom (Mo, W, etc.) and **X** is a chalcogen (S, Se, Te, etc.). TMDC possess non-zero bandgaps, which can be changed from indirect to direct ones depending on the number of layers, thus allowing many interesting application for electronic and optoelectronic devices [5-8]. Furthermore, their monolayer forms exhibit unique functionalities associated with electronic and spin degrees of freedom, thereby providing novel device concepts beyond conventional silicon-based devices [1-3,5-8]. Among TMDC, selenides have recently attracted a lot of attention due to their extraordinary optical tunability, catalytic and functional properties [9-11]. WSe$_2$ and MoSe$_2$ heterostructures are of particular interest due to their semiconductor bandgaps, easy synthesis and high reactivity [12,13]. There are several recent papers investigating selenides material properties as a basis for functional applications [14-16]. In this work we have investigated the mechanical properties of WSe$_2$ and MoSe$_2$ (single and few-layers) structures through fully atomistic molecular dynamics (MD) simulations.

**THEORY AND SIMULATION DETAILS**

Fully atomistic MD simulations were carried out using the LAMMPS code [17]. The universal force field (UFF) [18,19] was used to describe non-bonded interactions between atoms belonging to different layers of WSe$_2$/MoSe$_2$, as well as, silicon substrate layer interactions. For bonded interactions within each WSe$_2$/MoSe$_2$ layer the Stilling–Weber potential was used [19], parameterized from *ab initio* calculations [20]. The area of the layer considered for the calculations were 5 × 5 nm$^2$ and the size independence results was determined by calculating the force per atom over three different WSe$_2$/MoSe$_2$ cell sizes. The systems (WSe2 or MoSe2) consisted of different number of layers, as shown in Figure 1. The initial configurations were first optimized by a steepest-descent algorithm and then thermalized using a NVT ensemble (constant number of particles, volume and temperature) for 300 ps with a Nosé–Hoover thermostat [21,22] and time steps of 1 fs. After thermalization, two types of simulations were carried.

For the sliding simulations, a harmonic potential was applied to the top most layer. This was made by creating a spring-like between the layer and a reference point to provide the pulling direction. The used spring constant was 1 eV Å$^{-2}$ and the sliding velocity was set to 1 Å ps$^{-1}$. Using this approach, we determined the force required to start the sliding movement. For the stretching simulations, the box was increased along one determined direction on the layer surface, with all the atoms having their atomic positions updated. The strain rate was 0.002 Å fs$^{-1}$.

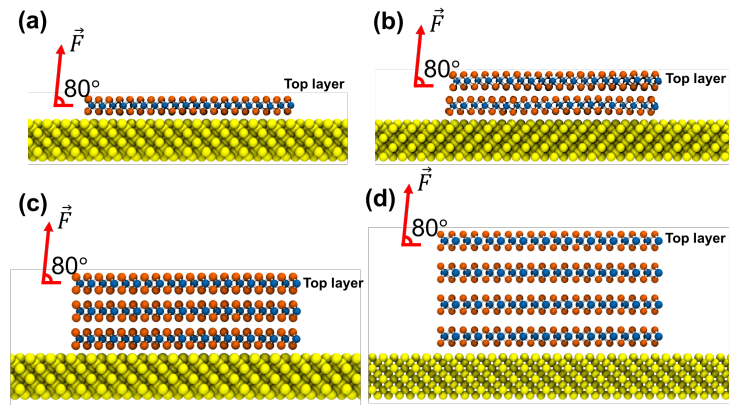

**Figure 1.** Initial configuration systems. See text for discussions.

**RESULTS AND DISCUSSION**

From peeling off experiments of single and few-layer WSe$_2$ and MoSe$_2$ structures deposited on silicon substrates [23], it was observed that single layer WSe2/MoSe2 have larger friction coefficients than 2, 3 and 4 layered structures. In order to understand this behavior from an atomistic point of view, we build structural models consisting of 1, 2, 3 and 4 WSe$_2$/MoSe$_2$ layers, as shown in Figure 1(a)–(d).

A force at 80° angle (in relation to basal layers) along the positive x direction is then applied to the topmost WSe$_2$ (or MoSe$_2$) layer in each setup configuration. This direction was chosen to provide a more realistic effect of the probe that was used in the peeling off experiments [23]. The external force is continuously increased until the topmost layer detaches from the rest of the system (layers and/or substrate). After equilibration, we observed that the adherence among layers and/or layers and substrate was robust, with very little movement of the flakes. The top layers of each system were then peeled off as shown in Figure 2 for the cases of 1 and 4 layers.

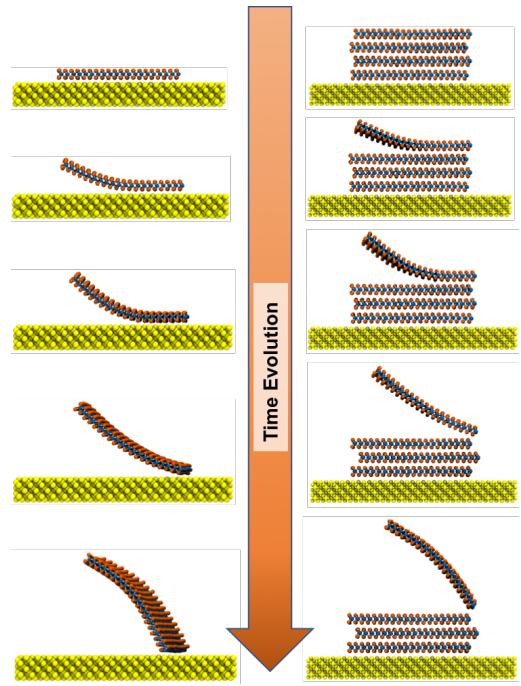

**Figure 2**. MD snapshots of the peeling off processes at different stages, for 1 (left) and 4 layers (right).

The layer detachment patterns are very similar, regardless of the number of layers and for all cases it is easier to peel off MoSe$_2$ structures. The main difference resides in the force value required to induce the detachment in each case, as shown in Figure 3(a). As we can see from this Figure, the magnitude of forces required to detach the first layer from the substrate is larger than 2nd, 3rd and 4th respectively. This is consistent with the available experimental data [23].

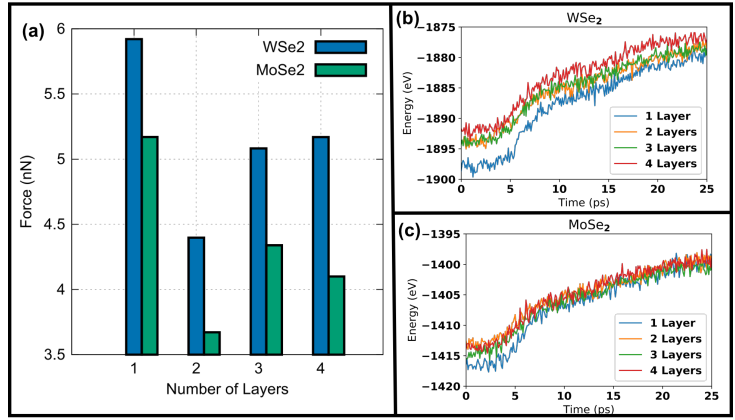

**Figure 3**. (a) Bar plot of force required to peel off one layer of WSe2/MoSe2 in the configurations shown in Figure 1. Potential energy as a function of simulation time of the monolayers being peeled off from silicon substrate (b) WSe2; (c) MoSe2.

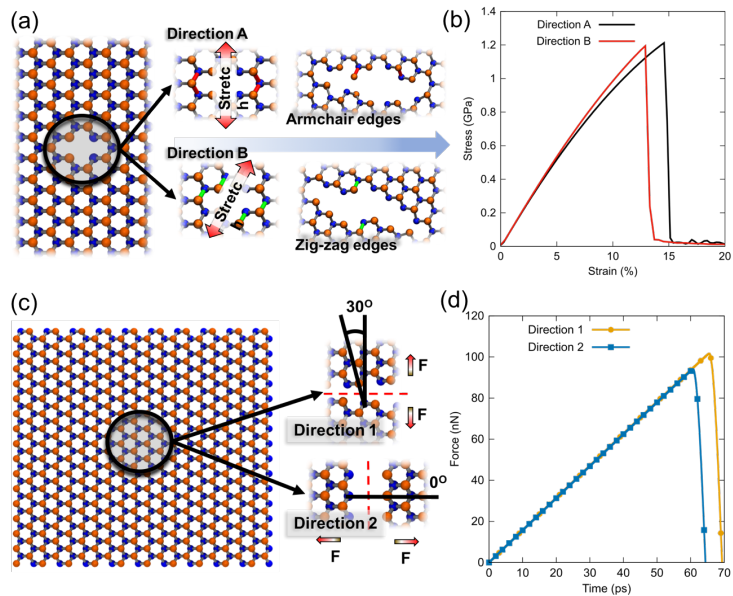

**Figure 4**. (a) Scheme of the defect/crack propagation, with armchair and zig-zag edges. (b) Stress versus strain curves (directions A and B). (c) Scheme of the induced fracture. Directions 1 and 2 refer to the applied force at a 30° and 0° angle with relation to W–Se bond, respectively. (d) Force as a function of simulation time (directions 1 and 2) for the fracture processes.

This can also be analyzed from the calculated potential energy of the topmost layer on the different scenarios (Figure 3(b,c)). As we can see from this Figure a larger energy value is necessary to peel off the monolayer cases in comparison to all multilayer configurations. It should be stressed that as the interactions among the layers are mainly van der Waals ones (described by a 6–12 Lennard–Jones-type potential), they rapidly decrease as the distance between pair of atoms increases. These results suggest that the interaction between 1st layer and the substrate is stronger in comparison to interlayer interactions themselves. Other 2D nanomaterials have also showed similar behavior and it has been speculated whether this is a universal-like behavior for layered materials [17].

With relation to multilayer systems, the bilayer configuration requires the smallest force to detach the topmost $WSe_2/MoSe_2$ layer. In this case, the topmost $WSe_2/MoSe_2$ is already distant from the substrate, thus minimizing their interactions. Now, the main interactions of the topmost layer are with the same material layer below it. Even though these $WSe_2/MoSe_2$ interlayer interactions are smaller in comparison to $WSe_2/MoSe_2$ -substrate ones, the perfect lattice match between $WSe_2/MoSe_2$ layers generates a well-stacked configuration. For 3 and 4 layers systems, we notice that this stacking is still important as the smaller distance between the $WSe_2/MoSe_2$ layers increases the interlayer interactions. This explains the differences on the force values for the systems with 2, 3 and 4 layers.

Another experimentally observed behavior was that when scratching/fracturing a monolayer, the required force is direction dependent [23]. In order to address this issue, we carried out further MD simulations creating a small defect on $WSe_2/MoSe_2$ layers, through removal of one W/Mo atoms and two Se ones. In order to investigate how crack propagates in these structures, we stretched these defected layers along two determined directions, as shown in Figure 4(a). Direction A generates high stress values on the set of bonds (depicted in red in Figure 4(a)), thus causing the opening of the crack with armchair edges. Direction B generates high stress values on another set of bonds (green ones in Figure 4(a)), which causes the opening of the crack with zig-zag-like edges. The calculated stress versus strain diagram for $WSe_2$ sheets along these two directions are shown in Figure 4(b). Even though the stress required to break the layers with zig-zag or armchair-like edges are similar, the one that forms zig-zag edges (Direction B) required less strain to open and split the structures.

Looking into pristine (no defects) $WSe_2$ layers (Figure 4(c)), we can better understand this behavior. We applied opposite forces on two groups of atoms within the layer, splitting the structure into half. Two directions were considered, Direction 1 and Direction 2, for splitting. Direction 1 makes an angle of 30° with W–Se bonds that are being stretched (breaking of these bonds forms armchair-like edges). On the other hand, Direction 2 is parallel to W–Se bonds being stretched, forming zig-zag-like edges. Figure 4(d) shows the force required to break the structure. Direction 1 showed larger force values required to break the structure.

**CONCLUSION**

We have investigated through fully atomistic MD simulations the mechanical properties of few-layer (from one up to four layers) of $WSe_2/MoSe_2$ deposited on a silicon substrate. Our results showed that single layer $WSe_2/MoSe_2$ have larger friction coefficients than 2, 3 and 4 layered structures. These results suggest that the interaction between 1st layer and the substrate is stronger in comparison to interlayer interactions themselves. Also, the scratching/fracture of the monolayers seems to be chirality dependent, with the preferential direction to crack propagation being the one perpendicular to a W(Mo)–Se bonds and zig-zag defects propagate much faster than armchair ones. $WSe_2$ and $MoSe_2$ structures show similar peeling off and scratch/fracture patterns, the main differences are in the force values involved in these processes. For all cases considered here it is always easier to peel off and/or fracture $MoSe_2$ structures.

ACKNOWLEDGMENTS

The authors would like to thank Dr. Ryan Major and Richard Nay of Hysitron for technical support. The authors thank the Air Force Office of Scientific Research (Grant FA9550-13-1-0084) for funding this research, and Air Force Office of Scientific Research MURI Grant FA9550-12-1-0035 financial support of this research. YMJ thanks São Paulo Research Foundation (FAPESP) Grant No. 2016/12341-5 for financial support. CFW thanks São Paulo Research Foundation (FAPESP) Grant No. 2016/12340-9 for financial support. YMJ, CFW and DSG acknowledge the Center for Computational Engineering and Sciences at State University of Campinas (FAPESP/CEPID grant No. 2013/08293-7).